\documentclass[twocolumn,prb,superscriptaddress,showpacs,preprintnumbers]{revtex4}

\usepackage{amsmath,dcolumn,graphicx}

\begin{document}


\title{Is U$_3$Ni$_3$Sn$_4$ best described as near a quantum critical point?}

\author{C. H. Booth} \affiliation{Chemical Sciences Division, Lawrence
Berkeley National Laboratory, Berkeley, California 94720, USA}

\author{L. Shlyk} \affiliation{
IFW, Institute for Solid State and Materials Research, POB 270016, 01171
Dresden, Germany}

\author{K. Nenkov} \affiliation{
IFW, Institute for Solid State and Materials Research, POB 270016, 01171
Dresden, Germany}
\affiliation{
International Laboratory of High Magnetic Fields and Low Temperatures,
Gajowicka 95, 53-529, Wroclaw, Poland}

\author{J. G. Huber} \affiliation{Department of Physics and Astronomy,
University of Kentucky, Lexington, KY 40506-0055, USA}

\author{L. E. De Long} \affiliation{Department of Physics and Astronomy,
University of Kentucky, Lexington, KY 40506-0055, USA}

\date{Submitted to PRB 4/17/3, Revised version 2.2 in press with PRB 22 January 2004}

\preprint{LBNL-52457}

\begin{abstract}

Although most known non-Fermi liquid (NFL) materials are structurally or chemically
disordered, the role of this disorder remains unclear.  In particular, very few 
systems have been discovered that may be stoichiometric and well
ordered.  To test whether U$_3$Ni$_3$Sn$_4$ belongs in this latter class, we
present measurements of the x-ray absorption fine structure (XAFS) of 
polycrystalline and single-crystal U$_3$Ni$_3$Sn$_4$ samples that are 
consistent with no measurable local structural disorder.  We also present 
temperature-dependent specific heat data in 
applied magnetic fields as high as 8~T that show features that are 
inconsistent with the antiferromagnetic Griffiths' phase model, but
do support the conclusion that a Fermi liquid/NFL crossover temperature 
increases
with applied field.  These results are inconsistent with
theoretical explanations that require strong disorder effects, but do
support the view that U$_3$Ni$_3$Sn$_4$ is a
stoichiometric, ordered material that exhibits NFL behavior,
and is best described as being near an antiferromagnetic quantum critical 
point.

\end{abstract}



\maketitle

\section{Introduction}

U$_3$Ni$_3$Sn$_4$ displays characteristic non-Fermi liquid behavior 
(NFL).\cite{Shlyk99}  For 
instance, the low-temperature magnetic susceptibility $\chi$ diverges 
as $T^{-0.3}$, 
the leading coefficient for the electronic term in the specific heat
$\gamma=C_\textrm{el}/T$ varies as $-T^{0.5}$, and the 
resistivity varies as $T^{1.79}$.  Such behavior is at odds with the standard
Fermi liquid (FL) description ($\chi \sim C_\textrm{el}/T \sim$ const., 
$\Delta \rho \sim T^2$) of Landau.\cite{Landau57} 
The present study attempts to differentiate between the applicability 
of various 
theoretical models describing NFL behavior in this system
by searching for the
presence of local lattice disorder and measuring how the electronic part of the specific 
heat is affected by applied magnetic fields.

Current models describing NFL behavior fall into a few general 
classes, including those that invoke close proximity to a zero-temperature phase transition,
competition between interactions such as Ruderman-Kittel-Kasuya-Yosida (RKKY) 
and Kondo effects, and those 
that include
magnetic-interaction disorder.  
For instance,
non-Fermi liquid behavior in the high-temperature
superconductors and in some of the heavy fermion systems has been postulated
to be due to the proximity of these systems to a zero-temperature magnetic
phase transition.\cite{NFL_mat_ex}  We will refer to such models
as anti- or ferromagnetic quantum critical point (AF-QCP or FM-QCP) models.
Indeed, a number of $f$-electron compounds and alloys 
have been described as near a QCP, based on the entry of a
system into a magnetic phase with increasing applied pressure 
or via chemical substitution (``chemical pressure'').\cite{Stewart01}  
Millis\cite{Millis93} and others\cite{Ioffe95,Moriya95,Lonzarich97,Coleman99} 
have developed the theory of critical fluctuations at 
temperatures above such a magnetic/non-magnetic QCP, 
building off earlier work by Hertz.\cite{Hertz76} The proximity of a 
magnetic phase is not necessary to obtain an NFL state, however.  For instance,
a multichannel Kondo model\cite{Nozieres80} also exhibits non-Fermi liquid 
properties.
In addition, even though these theories all use a single magnetic interaction 
strength between the $f$ and the conducting electrons (as opposed to a distribution), the first-discovered
NFL systems 
also contain some form of lattice disorder, usually in the 
form of chemical substitution.  This raises the possibility that lattice disorder
plays an important role in NFL physics.  In fact, a broad distribution of 
effective moments 
has been observed in  several systems (for instance, CeRhRuSi$_2$\cite{Graf97} 
and UPdCu$_4$\cite{Bernal95}).  These facts prompted researchers to 
consider the role of ``magnetic interaction disorder'' as a microscopic origin
for non-Fermi liquid effects.  One simple theory utilizes only Fermi liquid
concepts with a distribution of Kondo interactions, and is known as the ``Kondo
disorder model'' (KDM).\cite{Bernal95,Miranda96}  Other models that consider 
disorder
in the vicinity of a zero-temperature fixed point are known as Griffiths' phase
models (note that the KDM is also a Griffiths' phase model, but is not usually
classified as such).  
These include the Griffiths'-McCoy singularities that occur in a 
disordered Kondo system, but whose properties mainly derive from local
antiferromagnetic RKKY interactions within only a few 
clusters.\cite{Castro-Neto98,Castro-Neto00}  Another possible origin of a
Griffiths' phase occurs when disorder-induced Anderson localization occurs in 
the vicinity of a metal-insulator transition.\cite{Miranda01}  To clarify the
discussion, we will refer to the former model as the antiferromagnetic
Griffiths' phase, or AF-GP, and to the latter model as the 
metal-insulator-transition Griffiths' phase, or MIT-GP.

Although most known NFL materials have some intrinsic disorder, 
a few have recently been identified 
that appear to be stoichiometric and structurally well-ordered at 
ambient pressure.  Some examples include 
YbRh$_2$Si$_2$,\cite{Trovarelli00} CeNi$_2$Ge$_2$,\cite{Grosche00}
CeCoIn$_5$\cite{Petrovic01} and U$_3$Ni$_3$Sn$_4$.\cite{Shlyk99}  Although some
of the physical properties of these systems agree with those predicted by
the QCP model proposed by Millis,\cite{Millis93} none of these materials display 
properties that completely agree with it.  In addition, it is difficult to uniquely
differentiate between ``pure" QCP models and 
Griffiths' phase models, especially since the Griffiths' models have critical 
exponents that depend upon the degree of disorder in a way that is presently impossible to 
quantitatively relate to experimental measures of disorder.\cite{Castro-Neto_priv}

Careful consideration of the disorder-based and the pure QCP models involves 
comparisons both of electronic and magnetic properties to 
theory, and
thorough characterization of the degree of structural and magnetic order of the 
samples.  
The U$_3$Ni$_3$Sn$_4$ situation is complicated by the fact that
U$_3$Ni$_3$Sn$_4$ has been shown to have a Fermi liquid ground 
state below about 0.4-0.5 K, with NFL behavior occurring above this crossover
region.\cite{Shlyk00}  Regardless, in the NFL region, the electrical resistivity
goes as $\Delta \rho = \rho(T) - \rho_0 \sim T^{1.79}$.\cite{Shlyk99} 
This dependence is roughly consistent with the AF-QCP result of 
$\Delta \rho \sim T^{1.5}$, if one allows for the possibility, as discussed in 
Ref. \onlinecite{Coleman99}, that the non-Fermi liquid regions of the Fermi 
surface only occupy so-called ``hot lines" where magnetic scattering dominates, 
and that the rest of the Fermi surface constitutes a Fermi liquid regime that 
could dominate the conductivity.
The experimental magnetic susceptibility $\chi(T)$ diverges as $T^{-0.3}$,
although the lowest measured temperature is 2 K.  
AF-QCP systems should vary as $\chi(T)^{-1}=\chi_0^{-1}+cT^\alpha$, with 
$\alpha<1$, such as in the case of CeCu$_{5.9}$Au$_{0.1}$.\cite{Schroeder98}  
The U$_3$Ni$_3$Sn$_4$ data can also be fit with this form, although the 
accuracy of the final result ($\alpha=0.3 \pm 0.2$) is limited by the measured 
temperature range.  In any case, the magnetic susceptibility data can also be 
interpreted as consistent with the AF-QCP.
The electronic part of the specific heat is also consistent with the AF-QCP,
varying as as $C_\textrm{el}/T \sim -T^{0.5}$.  Alternatively,
these results can be self-consistently explained with an AF-GP 
phase\cite{Castro-Neto98,Castro-Neto00} and a critical exponent of
$\lambda=0.7$, which produces comparably good fits to the 
data.\cite{Shlyk00}  In addition, these data are qualitatively consistent with 
the two-channel Kondo model\cite{Cox87,Cox98,Jarrell96,Jarrell97}, although fits
using this model require an unrealistically high spin-fluctuation 
energy.\cite{Shlyk99} Comparisons to the KDM are not favorable either, 
since the KDM predicts logarithmic divergences of the magnetic susceptibility 
and specific heat and a linear temperature dependence of the electrical
resistivity, all of which are clearly at odds with the experimental data.
In addition, measurements at applied pressures up to 1.8 GPa\cite{Estrela01}
indicate that the low-temperature FL ground state of U$_3$Ni$_3$Sn$_4$ 
extends to higher temperatures with increasing
pressure.  A scaling analysis of the FL/NFL crossover temperature as a function
of applied pressure strongly implies a magnetic critical point at a 
negative-pressure that has been estimated at $-0.04\pm0.04$ GPa.  

Although the measured properties of U$_3$Ni$_3$Sn$_4$ do not clearly
support any of the various NFL models, there
is little evidence to suggest that any disorder exists in this system, 
which is inconsistent with both the KDM and Griffiths' phase models in spite of 
their agreement with thermal and magnetic data.  In particular, single crystals 
of the material form, and x-ray diffraction studies of available crystals show 
the material to be consistent with the nominal
stoichiometry.  Moreover, the residual resistivity is as low as
7 $\mu\Omega\cdot$cm\cite{Shlyk99}.  
Nevertheless, some forms of lattice disorder 
can be difficult to detect using standard diffraction techniques.  For instance,
if disorder occurs in a non-periodic fashion, such as in amorphous regions or
very small domains, only a local structural probe such as x-ray absorption
fine-structure (XAFS) or pair-distribution function (PDF) analysis of powder
diffraction data will be sensitive to it.  In addition, 
no temperature dependent structural studies have been performed,
and disorder broadening of the mean-squared displacement parameters 
(Debye-Waller factors) can easily be confused with
large vibrational amplitudes.
Therefore, we have undertaken a direct test of the degree of structural order in
U$_3$Ni$_3$Sn$_4$ using XAFS 
spectroscopy. 
Although some technical issues limit the accuracy of the estimated
maximal disorder levels as detailed below, our measurements are consistent with
no disorder within experimental error, for both single crystals and
polycrystals, based on temperature dependent data from all three investigated absorption edges.

As implied above, merely having the lattice be (measurably) well ordered may not
rule out magnetic interaction disorder.  
Since the Griffiths' phase models only require very few clusters to be 
dominated by one of the competing interactions, such clusters may only have a 
negligible effect on the average lattice disorder in a real material.  In
addition, some heretofore more subtle Kondo disorder mechanism may still be 
applicable (for instance, if large fluctuations in the conduction electron
density of states accompany fluctuations in the hybridization strength in 
the presence of lattice disorder\cite{Cox_priv}).  In any case, the AF-GP model 
makes quantitative
predictions regarding the magnetic field dependence of certain properties, 
including the specific heat.  Below we compare such measurements to the AF-GP 
predictions and find they are inconsistent.  Instead, we find these data are
more consistent with the presence of a low-lying magnetic phase.  

The rest of this paper is organized as follows.  The XAFS measurements are
described in Sec. \ref{XAFS}, including a description of the sample preparation
and the XAFS technique.  Sec. \ref{CHT} relates the results from the measurements of
specific heat as a function of temperature and applied field.
All of these results are discussed in relation to the various NFL theories in 
Sec. \ref{discussion}, and the final conclusions are summarized in 
Sec. \ref{conclusion}.

\begin{figure}[t]
\includegraphics[width=2.5in, angle=90,clip]{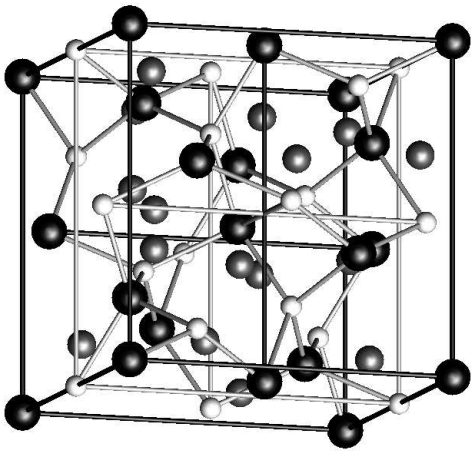}
\caption{Crystal structure of U$_3$Ni$_3$Sn$_4$.  Black balls represent
uranium, gray tin and white nickel.  The material is \textit{bcc} with
space group $I\bar{4}3d$ and $a$=9.3524 \AA.
}
\label{xtal}
\end{figure}

\begin{figure}[b]
\includegraphics[width=3.4in, trim=50 0 50 110,clip]{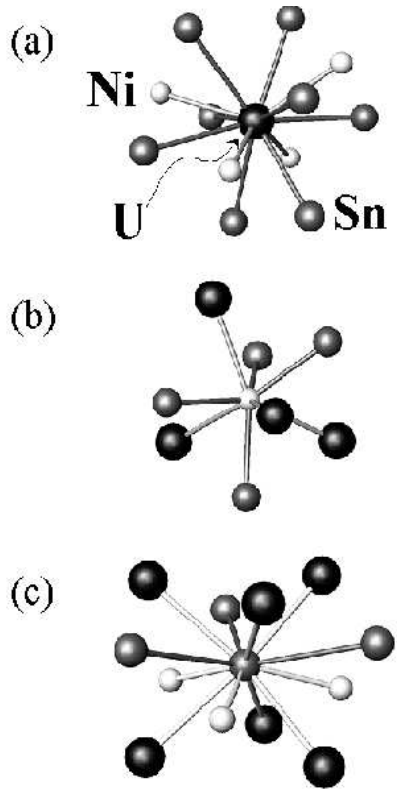}
\caption{Near-neighbor coordinations of the (a) uranium, (b) nickel, and
(c) tin sites.  See Sec. \ref{xafs_bg} for details.
}
\label{local_xtal}
\end{figure}

\begin{figure}[t]
\includegraphics[width=3.4in,trim=0 22 0 0,clip]{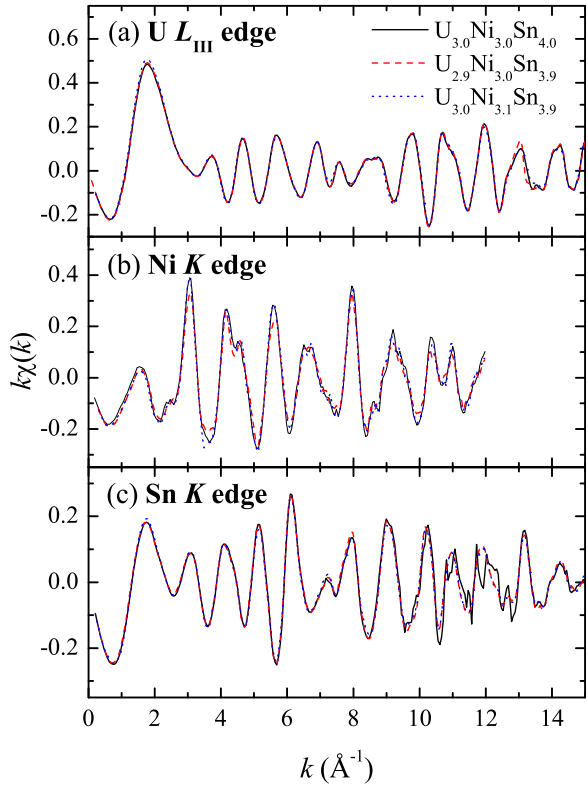}
\caption{XAFS data for the three polycrystalline 
samples.  Single-crystal data are quantitatively similar.  Data from the
various samples are very similar, and so are difficult to differentiate in the 
plot.
}
\label{poly_ks}
\end{figure}

\begin{figure}[t]
\includegraphics[width=3.4in,trim=0 22 0 0,clip]{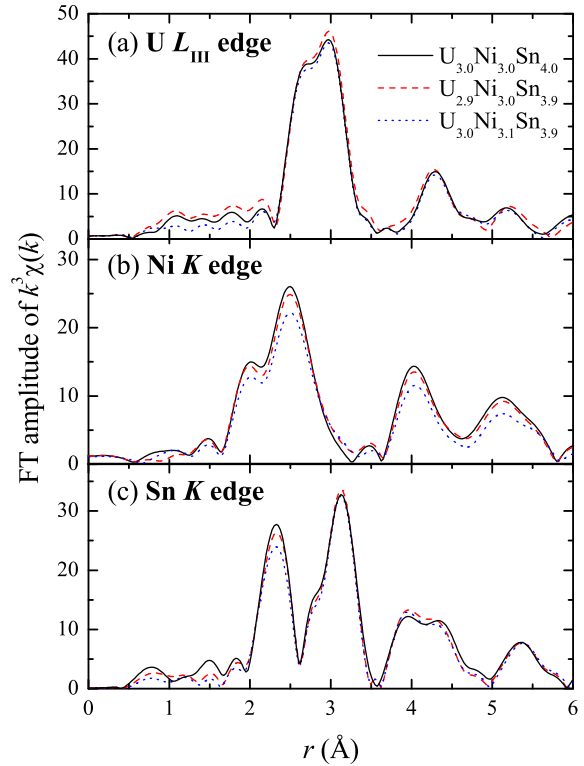}
\caption{Fourier transforms of the $k^3\chi(k)$ XAFS data for the 
polycrystalline samples.  
U and Sn (transmission) transforms are from $k$=3.0-15 \AA$^{-1}$, 
while the Ni (fluorescence) transform is from 2.5-12.0 \AA$^{-1}$, and all 
transform windows are Gaussian narrowed by 0.3 \AA$^{-1}$.
}
\label{poly_rs}
\end{figure}

\begin{figure}[t]
\includegraphics[width=3.4in,trim=0 22 0 0,clip]{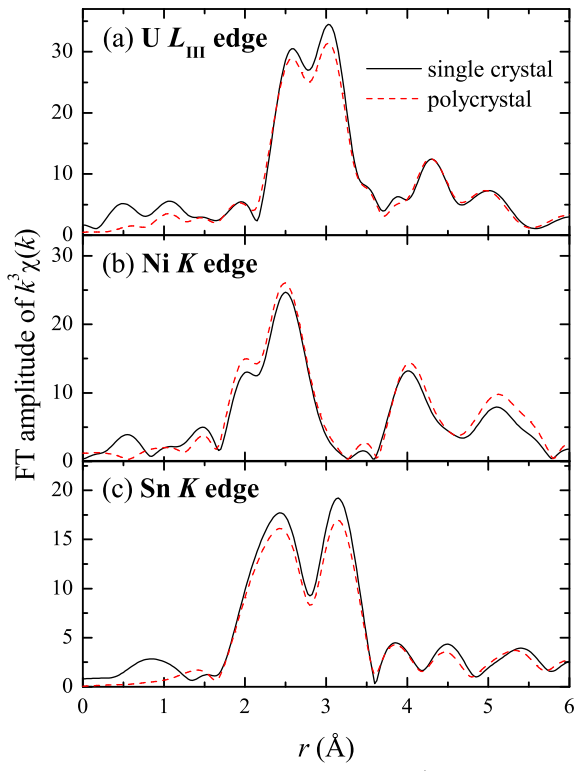}
\caption{Fourier transforms of the $k^3\chi(k)$ XAFS data for the 
nominally U$_3$Ni$_3$Sn$_4$ single crystal.  Data for the nominally 
U$_3$Ni$_3$Sn$_4$ polycrystalline sample with the same transform ranges 
are shown for comparison.  U and Sn transforms are from $k$=3.0-13 \AA$^{-1}$
while the Ni transform is from 2.5-12.0 \AA$^{-1}$, all
Gaussian narrowed by 0.3 \AA$^{-1}$.
}
\label{xtal_poly_rs}
\end{figure}

\begin{figure}[t]
\includegraphics[width=3.4in,trim=0 22 0 0,clip]{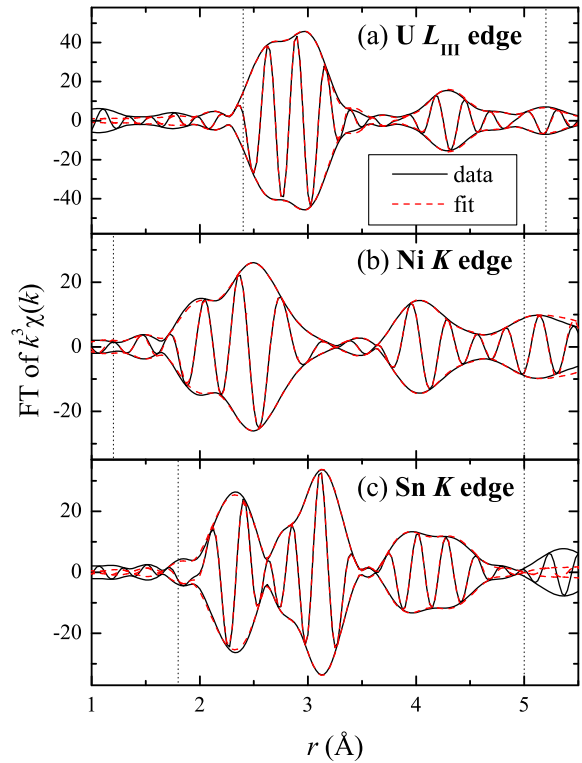}
\caption{Examples of the fits to the (a) U $L_\textrm{III}$- and the
(b) Sn $K$-edge polycrystalline data.  Each transform is represented by three lines.  
The inner oscillating line is the real part of the complex transform, while the 
envelope lines are $\pm$ the amplitude of the transform.
Vertical dotted lines show the $r$-space fit range.
Transform ranges are as in Fig. \ref{poly_rs}.  
}
\label{rs_fit}
\end{figure}

\begin{figure}[t]
\includegraphics[width=3.4in,trim=0 42 0 0,clip]{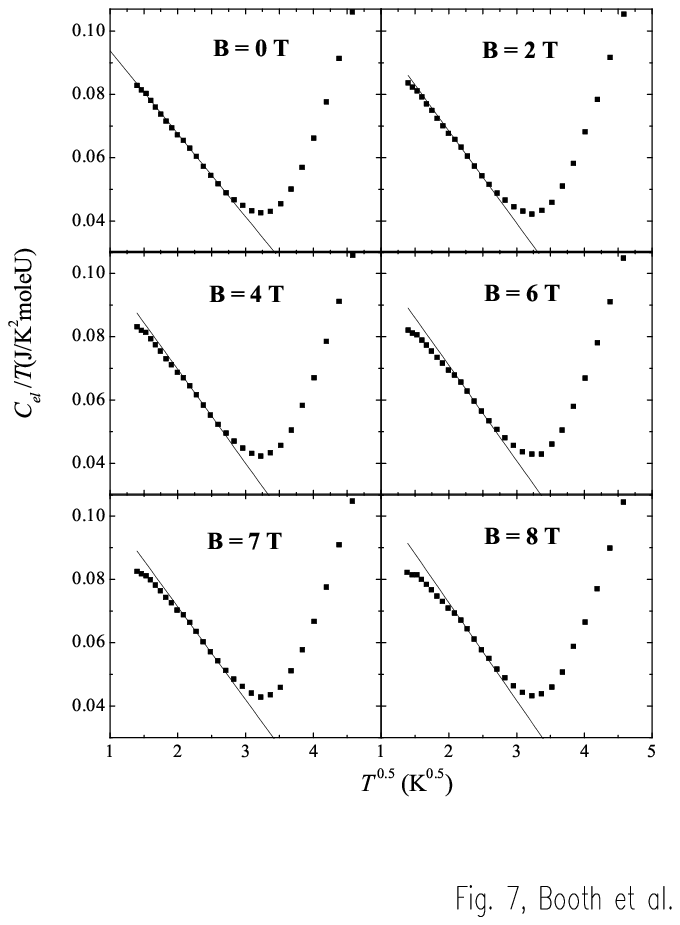}
\caption{
$C_\textrm{el}/T$ as a function of $T^{0.5}$ of U$_3$Ni$_3$Sn$_4$ single crystal for 
$B$ = 0, 2, 4, 6, 7, 8~T. The straight line is a guide to the eye for the 
$C_\textrm{el}/T \sim T^{0.5}$ behavior.
}
\label{spec_fig}
\end{figure}

\begin{figure}[t]
\includegraphics[width=3.4in,trim=0 30 0 0,clip]{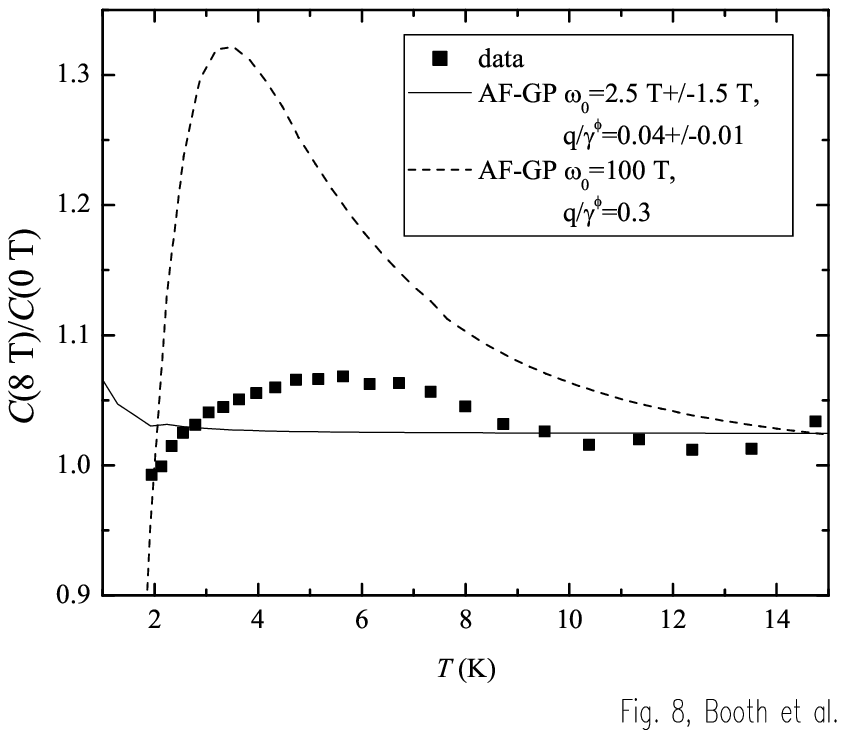}
\caption{
The ratio of $C(H=8 \textrm{T},T)/C(H=0 \textrm{T},T)$ from Fig. \ref{spec_fig}.
Lines are fits to the AF-GP model described in the text.
}
\label{AFGP_fig}
\end{figure}

\section{XAFS measurements}
\label{XAFS}

\subsection{Background}
\label{xafs_bg}

U$_{3}$Ni$_{3}$Sn$_{4}$ crystallizes into a $bcc$ structure, in the $I\bar{4}3d$
space group with the room-temperature lattice parameter 9.3524 \AA~and a 
position parameter $x$=0.082
describing the Sn (16$e$) site (Fig. \ref{xtal}).\cite{u334_param}  
The near-neighbor shells are fairly well separated (Fig. \ref{local_xtal}).  
For instance,
U has 4 nearest-neighbor Ni's at 2.86 \AA, followed by 8 Sn neighbors at 
3.24 \AA.  Ni has 4 Sn neighbors at 2.61 \AA~and 4 U neighbors at 2.86 \AA.  
Sn has 3 Ni neighbors at 2.61 \AA,
followed by 6 U neighbors at 3.24 \AA, and 3 Sn neighbors at 3.50 \AA.

\subsection{Experimental}

Three of the samples are polycrystalline with nominal stoichiometries 
U$_{3.0}$Ni$_{3.0}$Sn$_{4.0}$, U$_{2.9}$Ni$_{3.0}$Sn$_{3.9}$, and
U$_{3.0}$Ni$_{3.1}$Sn$_{3.9}$. Two single-crystal samples were also measured
with nominal stoichiometries of U$_{3.0}$Ni$_{3.0}$Sn$_{4.0}$ and 
U$_{2.9}$Ni$_{3.0}$Sn$_{3.9}$.  The stoichiometries of the single crystals has
been confirmed by single-crystal x-ray diffraction measurements and are,
in fact, the same samples as those reported in Ref. \onlinecite{Shlyk99}.
Polycrystalline sample stoichiometries are only nominally based on the 
composition of the starting materials.  However, we believe they are accurate 
given that the sample weight losses from arc melting were always the order of a 
few tenths of a percent or less, and could be entirely attributed to a tendency 
for sample boules to spall on the surface or shatter when first struck with the 
arc discharge.  Uranium has an extremely low vapor pressure during arc melting, 
and noteworthy weight losses of Sn or Ni material were not observed at any 
composition near the desired stoichiometry.

All XAFS data were collected on beam lines 4-1 and 4-3 at the Stanford 
Synchrotron Radiation Laboratory (SSRL) using half-tuned Si(220) double 
monochromator crystals.  Samples were placed into a LHe flow cryostat.  
Polycrystalline samples were ground into a fine powder under acetone, passed 
through a 40 $\mu$m sieve and brushed onto scotch tape, with stacked layers 
such that the total thickness of each transmission sample corresponded to a 
change of about one absorption length at each measured edge.  Data for the 
polycrystals were collected at various temperatures between 20~K and 300~K at 
the U $L_{\textrm{III}}$ and Sn $K$ edges in transmission mode, and at the Ni 
$K$ edge in fluorescence mode using a 4-pixel Ge detector.\cite{Bucher96}  
Single crystal data were collected at 
20~K at the U $L_{\textrm{III}}$ and the Sn and Ni $K$ edges in fluorescence
mode.  Several scans were obtained
for each sample at each edge and temperature, and were fit separately to 
crosscheck the error estimates.  Dead-time and 
self-absorption\cite{Booth_thesis}
corrections were applied to the fluorescence data.  

Data were reduced and fit in position-space
using the RSXAP package.\cite{Hayes82,Li95b,RSXAP}  In particular, the XAFS
function $\chi(k)$ is defined as $\mu(k)/\mu_E(k)-1$, where $\mu(k)$ is the 
absorption coefficient as a function of the photoelectron wave vector $k$, 
and $\mu_E(k)$ is the so-called ``embedded atom'' background absorption that
is proportional to the number of generated photoelectrons.  The wave vector
is defined as $k=\sqrt{\frac{2 m_e}{\hbar^2}(E-E_0)}$. The photoelectron
threshold energy $E_0$ is defined arbitrarily as the energy at the half-height of the 
edge, and is allowed to vary in subsequent fits.
Examples of the
$k\chi(k)$ data are shown in Fig. \ref{poly_ks} for the polycrystalline samples.
Data on single crystals are similar, both in quality and quantity.

The scattering amplitudes are all fixed to $N_i S_0^2$, where $N_i$ is
the nominal number of neighbors in the $i$th shell for the stoichiometric 
compound, and $S_0^2$ is the XAFS amplitude scale factor.
Each data set was fit with a single value of 
$S_0^2$, assuming full nominal site occupancies.  
All scattering paths also share
a single value of $E_0$.  In the case where multiple temperatures were 
collected, average values of $S_0^2$ and $E_0$ were obtained and then held fixed for
all temperatures for a given edge.  
Fixing $N_i$, $S_0^2$ and $E_0$ in the final fits drastically reduces the
number of fit parameters, but assigns all lattice disorder effects to 
either the measured bond lengths, $R_i$, or the Debye-Waller factors, $\sigma_i$.
In particular, the effect of vacancies is placed on the Debye-Waller factors.
Measured $S_0^2$ and $E_0$ values for both the single-crystal and the polycrystalline data are the same within experimental error.
Reported error estimates use the
larger of either a 
Monte Carlo estimate of the 1-standard deviation displacements 
(essentially equivalent to a 
covariance matrix without having to assume that the statistical-$\chi^2$ is
quadratic near its minimum), or the width of the distribution of parameters 
obtained by fitting the individual scans at each temperature.  Reported errors 
are generally consistent with those obtained from comparisons to standard 
compounds, typically $\pm$0.005~\AA~in pair distance and $\pm$10-20\% in 
$\sigma^2$ for near neighbors, with the error roughly doubling after about 
3~\AA.\cite{Li95b}

\squeezetable
\begin{table*}
\caption{Final fit parameters to the U $L_{\textrm{III}}$-, Ni $K$- and Sn $K$- 
edge data 
at 20 K on three polycrystalline samples of U$_3$Ni$_3$Sn$_4$ with various nominal 
stoichiometries.  U $L_{\textrm{III}}$ edge fits have
$S_0^2 = 0.73 \pm 0.06$ and $\Delta E_0 = -10.3 \pm 0.4$.  Ni $K$ edge
fits are from 2.5 to 13.0 \AA$^{-1}$ (Gaussian narrowed by
0.3 \AA$^{-1}$) and from 1.4 to 5.0 \AA, and
have $S_0^2 = 0.64 \pm 0.04$ and $\Delta E_0 = -0.5 \pm 0.5$.
Sn $K$ edge fits have
$S_0^2 = 0.95 \pm 0.06$ and $\Delta E_0 = -8.3 \pm 0.1$. Diffraction data is 
provided for comparison ($R_{\textrm{diff}}$) and is
from Ref. \onlinecite{Shlyk99} on a single-crystal sample of U$_3$Ni$_3$Sn$_4$
collected at room temperature.
}
\begin{ruledtabular}
\begin{tabular}{lcccccccccccccc}
&&
&\multicolumn{4}{c}{U$_3$Ni$_3$Sn$_4$} 
&\multicolumn{4}{c}{U$_{2.9}$Ni$_{3.0}$Sn$_{3.9}$} 
&\multicolumn{4}{c}{U$_{3.0}$Ni$_{3.1}$Sn$_{3.9}$}
\\
pair & $N$ & $R_{\textrm{diff}}$ 
& $R$ & $\sigma^{2}$ 
& $\sigma^2_\textrm{static}$ & $\Theta_\textrm{cD}$ 
& $R$ & $\sigma^{2}$ 
& $\sigma^2_\textrm{static}$ & $\Theta_\textrm{cD}$ 
& $R$ & $\sigma^{2}$ 
& $\sigma^2_\textrm{static}$ & $\Theta_\textrm{cD}$ \\
\colrule
U-Ni & 4 & 2.864 
& 2.848(3) & 0.0018(2) & -0.0009(4) & 252(5) 
& 2.848(4) & 0.0019(4) & -0.0004(5) & 282(2) 
& 2.848(3) & 0.0019(2) & -0.0005(5) & 259(4) 
\\
U-Sn & 8 & 3.237 
& 3.228(2) & 0.0011(2) & -0.0006(3) & 241(1) 
& 3.226(4) & 0.0009(2) & -0.0009(3) & 231(1) 
& 3.226(3) & 0.0009(2) & -0.0007(3) & 233(1) 
\\
U-U  & 8 & 4.374 
& 4.36(1)  & 0.0016(3) & -0.0005(3) & 164(6) 
& 4.355(5) & 0.0014(2) & -0.0005(3) & 173(2) 
& 4.355(3) & 0.0014(2) & -0.0000(3) & 169(3) 
\\
U-Ni & 2 & 4.676 
& 4.67(1)  & 0.0015(3) &            &        
& 4.67(1)  & 0.0022(6) &            &        
& 4.67(1)  & 0.0022(4) &            &        
\\
\\
Sn-Ni& 3 & 2.609 
& 2.604(3) & 0.003(1)  &  0.001(1) & 420(15)
& 2.597(3) & 0.0027(2) & -0.0003(2) & 349(4) 
& 2.599(3) & 0.0027(2) &  0.0008(3) & 359(3) 
\\
Sn-U & 6 & 3.237 
& 3.223(5) & 0.0006(4) & -0.0017(5) & 202(12) 
& 3.232(7) & 0.0016(2) & -0.0004(2) & 246(2) 
& 3.228(3) & 0.0012(2) & -0.0004(2) & 273(5) 
\\
Sn-Sn& 3 & 3.497 
& 3.50(3) & 0.003(3)  & -0.003(3)  & 172(17)
& 3.500(4) & 0.004(1)  & -0.000(1)  & 245(6) 
& 3.496(3) & 0.0017(5) & -0.0001(6) & 250(20) 
\\
Sn-Sn& 2 & 4.050 
& 3.98(3)  & 0.003(3)  &            &        
& 4.03(1)  & 0.01(1)   &            &        
& 4.02(1)  & 0.0024(7) &            &        
\\
Sn-Ni& 3 & 4.232 
& 4.25(3)  & 0.002(1)  &            &        
& 4.16(5)  & 0.01(1)   &            &        
& 4.22(1)  & 0.005(2)  &            &        
\\
Sn-Sn& 6 & 4.594 
& 4.60(3)  & 0.002(1)  &            &        
& 4.598(4) & 0.0034(3) &            &        
& 4.596(3) & 0.0023(2) &            &        
\\
\\
Ni-Sn& 3 & 2.609 
& 2.588(4) & 0.0017(4) & -0.0004(5) & 347(10)
& 2.585(6) & 0.0014(3) & -0.0007(3) & 334(10)
& 2.587(4) & 0.0017(6) &  0.0003(6) & 353(4) 
\\
Ni-U & 3 & 2.864 
& 2.850(6) & 0.0028(6) &  0.0003(7) & 252(7)
& 2.850(8) & 0.0023(5) & -0.0005(5) & 234(7) 
& 2.853(6) & 0.002(1)  & -0.0005(9) & 234(4) 
\\
Ni-Sn& 3 & 4.231 
& 4.24(1)  & 0.0005(9) &            &       
& 4.26(1)  & 0.000(2)  &            &        
& 4.25(2)  & 0.004(3)  &            &        
\\
Ni-Ni& 3 & 4.374 
& 4.35(1) & 0.001(1)  &            &        
& 4.35(1) & 0.000(1) &            &        
& 4.37(1) & 0.001(2) &            &        
\\
Ni-U & 3 & 4.676 
& 4.70(5)  & 0.005(7)  &            &        
& 4.70(3)  & 0.003(4) &            &        
& 4.70(3)  & 0.003(2) &            &        
\end{tabular}
\end{ruledtabular}
\label{poly_table}
\end{table*}

\subsection{Results}

Figures \ref{poly_rs} and \ref{xtal_poly_rs} show the Fourier transforms (FT) 
of $k^3\chi(k)$.  Peaks in the FT's correspond to near-neighbor pairs 
distances in the local structure. Although these functions are closely
related to the radial distribution function, there are some important
differences.  For instance,
constructive/destructive interference can occur (the functions have real and 
imaginary parts), the scattering profiles are not Gaussian,
and phase shifts occur that place the peaks at distances in the FT's that are
shorter than in the structure by an amount related to the species of absorber
and backscatterer.  All of these complications are included in the detailed 
fits below. Transmission data were collected out to a
$k_{\textrm{max}} = 15$ \AA, and fluorescence data were collected out to a
$k_{\textrm{max}} = 13$ \AA.  Fig. \ref{xtal_poly_rs} shows a comparison
between the single-crystal and polycrystalline data for the 
U$_{3.0}$Ni$_{3.0}$Sn$_{4.0}$ samples using the same transform ranges.

There are visible differences between the various data sets for a given edge.  
In particular, the Ni and Sn $K$-edge data on the polycrystalline 
U$_{3.0}$Ni$_{3.1}$Sn$_{3.9}$ sample
consistently show a reduced amplitude compared to U$_{3.0}$Ni$_{3.0}$Sn$_{4.0}$
at all temperatures, consistent with some 
disorder or the presence of an amorphous phase containing those elements.
Although these differences are above the signal to noise, it is not possible 
to discern their exact cause 
from the fit results listed below.  Differences between
the polycrystalline and the single-crystal data (Fig. \ref{xtal_poly_rs})
are similar in magnitude, but complications with 
analyzing single-crystal data, such as dead-time and self-absorption 
corrections, are very likely the cause.  

The basic procedure used here for searching for lattice disorder is to carry out
fits assuming the nominal structure, then
examine certain parameters for signs of disorder.  In the fits, each distinct
scattering shell in the nominal structure out to about 4.7 \AA~is used
at each edge.  
Fit results to the data from the polycrystalline samples are reported in 
Table \ref{poly_table}, and are 
compared to results from diffraction measurements. The fit quality is
very high; examples are shown in Fig. \ref{rs_fit}.  
Although all the polycrystalline data were collected as a function of temperature,
we only show the fit results for the coldest measured temperature.  No
significant changes in the fit parameters occur with temperature, except that 
the Debye-Waller factors increase in a manner consistent with the 
correlated-Debye model\cite{Crozier88} plus a temperature-independent offset 
$\sigma^2_\textrm{static}$.  
Such offsets can be used as indicators for 
non-thermal disorder (a prime example occurs in the colossal magnetoresistance 
manganese perovskites\cite{Booth98a}). 
Results for 
the correlated-Debye temperature
$\Theta_\textrm{cD}$ and $\sigma^2_\textrm{static}$ are shown in Table 
\ref{poly_table}. 
We see no evidence for 
$\sigma^2_\textrm{static}$ values inconsistent with zero disorder. 
Note that the preponderance of small, negative values 
of $\sigma^2_\textrm{static}$ are unphysical and likely due to small 
underestimates of $S_0^2$.

Fit results to the single-crystal data were found to be the same as results
from the polycrystalline data within the error estimates, and so are not reported 
here.  
No evidence for lattice
disorder is observed, as exemplified by the consistently low values of the
nearest-neighbor $\sigma^2$'s for the low temperature fits.

Finally, we consider the possibility of site interchange, or site/anti-site disorder.  This possibility is very remote, however, given the big 
differences in the radii of the atoms involved, except for Sn/U interchange 
(covalent radii are 1.42 \AA, 1.15 \AA, and 
1.41 \AA~for U, Ni and Sn, respectively).  Unfortunately, fitting the Sn and U
XAFS data including some U/Sn interchange gives only a broad result:
$s = 9 \pm 10$\%, where $s$ is the percentage of Sn sitting on U (12$a$) sites.
Fits including U/Ni and Sn/Ni site interchange were similarly imprecise.
The principal difficulty in using XAFS (or diffraction, for that matter) to 
measure $s$ between two atomic species is the correlation between $s$ and the 
Debye-Waller factors for each site.  If $s$ is sufficiently large, some of that
uncertainty is removed.  This unfortunate situation is best illustrated by 
looking at the polycrystalline data in Fig. \ref{poly_rs}.  These various 
samples have different stoichiometries, and obvious systematic differences in
the XAFS Fourier transforms are visible that certainly are at least partially due to the
various "site interchanges."  However, they all fit a 334-type stoichiometry well
(Table \ref{poly_table}).  In any case, since there are no easily visible peaks 
corresponding
to, say, the Sn site in the U data (Figs. \ref{poly_rs}, \ref{xtal_poly_rs}), 
we conclude that there is probably even less site interchange than allowed for 
by the upper limits reported above.
This situation is in contrast to that in the
UPdCu$_4$ system.\cite{Booth98b}
In any case, the single-crystal diffraction results\cite{Shlyk99} should not have produced 
such high quality fits if much more than 5\% of such interchange
occurs.

\section{Field-dependent specific heat measurements}
\label{CHT}

\subsection{Experimental}

The heat capacity measurements were performed using a Physical Property 
Measurement System (Quantum Design).  The temperature was controlled by a 
Cernox thermometer. The temperature error is 1\% at 4 K and 9 T. The heat 
capacity software uses Quantum Designs' ``two-$\tau$'' model to measure the heat
capacity of the sample. The two-$\tau$ model simulates the effect of heat 
flowing between the sample platform and sample, and the effect of heat flowing 
between the sample platform and thermal bath.   

\subsection{Results}

The specific heat of a U$_3$Ni$_3$Sn$_4$ single crystal from the same batch as 
the XAFS sample was measured between 1.8 and 30~K in applied magnetic fields up 
to 8~T.  These data are shown in Fig.~\ref{spec_fig}, plotted as 
$C_\textrm{el}/T$ versus $T^{0.5}$.  Here we 
have already subtracted the hyperfine and lattice contribution according to 
the specific heat analysis reported previously.\cite{Shlyk00}
Note that although a nuclear Schottky term exists
for the Sn atoms in the sample, its effect can be seen to be negligible for this
analysis.\cite{Shlyk00} 
In zero applied magnetic field the data follow the 
$C_\textrm{el}/T \propto -T^{0.5}$ behavior below 6~K, indicative of the
non-Fermi liquid regime. Increasing the applied magnetic field progressively 
depresses the specific 
heat so that $C_\textrm{el}/T$ shows a deviation from square root behavior at 
lower temperatures. It is expected that $C_\textrm{el}/T$ tends toward a 
constant value at temperatures lower than 1.8~K, suggesting the onset of a 
Fermi-liquid regime. 
These results strongly imply that the applied fields destroy the magnetic
fluctuations due to a nearby antiferromagnetic critical point.  

On the other hand, if one plots the data as 
$C_\textrm{el}(H,T)/C_\textrm{el}(H=0,T)$ (Fig. \ref{AFGP_fig}), one might 
interpret the observed peak near 5~K in applied
field as arising from a Schottky-like feature.
This observation leads to an alternative explanation of these data
provided by the AF-GP model.\cite{Castro-Neto00}  In the high-field
limit of this model, the specific heat should go as 
$C_\textrm{el}/T = A(H^{2+\lambda/2}/T^{3-\lambda/2})e^{-\mu_\textrm{eff}H/T}$, 
where $\lambda=0.7$ is the aforementioned
critical exponent from the low temperature $C_\textrm{el}(H=0)/T$ and magnetic 
susceptibility data, $\mu_\textrm{eff}$ is an average effective moment of the 
antiferromagnetic clusters, and $A$ is a constant that is difficult to 
calculate in the theory and is thus taken as arbitrary.  This function 
has been successfully applied to, for instance,
Ce$_{0.05}$La$_{0.95}$RhIn$_5$.\cite{Kim02}   However, this
analysis leaves open the question of 
whether the coefficient $A$ takes on physically meaningful values, and whether
the aforementioned assumption that the data is in the high-field limit is valid.
Instead,
consider the form for the specific heat at any field given in 
Ref. \onlinecite{Castro-Neto00}:
\begin{eqnarray}
C_\textrm{el}(H,T)  \propto \beta^2 \int^{\omega_0}_{0} d\Delta \Delta^{1-\lambda}
(E_H^2+\Delta^2)
\nonumber \\*
\times\ \textrm{sech}^2(\beta\sqrt{E_H^2+\Delta^2}) 
\left[ \ln \frac{\omega_0}{\Delta} \right] ^{1-\theta},
\label{GPCH}
\end{eqnarray}
where $\Delta$ is the cluster
tunneling energy, $\omega_0$ is the tunneling energy for a single atom, $\beta$
is $1/k_\textrm{B}T$, and
$\theta$ is the percolation exponent.  $E_H$ is the magnetic energy of a
given cluster, and is given by:
\begin{equation}
E_H(\Delta)=q\mu_\textrm{B} \left[\frac{1}{\gamma}\ln\left(\frac{\omega_0}{\Delta}\right)\right]^\phi H,
\end{equation}
where $q$ gives the magnitude of the average moment within a cluster, and
$\phi=1(1/2)$ for ferromagnetic (antiferromagnetic) interactions.
For our data, $\lambda \approx 0.7$, the
percolation scaling exponent for three dimensions with no magnetic order is 
$\theta=3/2$, and $\phi=1/2$.  The tunneling frequency cutoff $\omega_0$
is taken as one of only two fitting variables, with the other being
$q/\gamma^\phi$, which we apply as a single variable.  By taking the ratio
$C_\textrm{el}(H=8T,T)/C_\textrm{el}(H=0,T)$ (valid at all fields and 
temperatures), we eliminate the coefficient $A$.  Using this form, we find that
no combination of fitting variables produces a satisfactory fit.  For example, 
Fig. \ref{AFGP_fig} shows a typical calculation where the parameters were 
chosen to give a peak in 
$C_\textrm{el}(H=8T,T)/C_\textrm{el}(H=0,T)$ in the vicinity of the observed 
peak.  This ``fit'' produces far too large a peak compared to the peak in the
data.  The best fit actually 
places the Schottky-like anomaly below the observed range with a very small 
effective moment ($\sim$1/100$^\textrm{th}$ of that from 
Ce$_{0.05}$La$_{0.95}$RhIn$_5$).  We therefore 
conclude that the AF-GP model (as posed) does not describe the physics in
U$_3$Ni$_3$Sn$_4$.

\section{Discussion}
\label{discussion}

Deviations from the nominal structure in the fit results can occur in a number
of ways.  First, the measured $S_0^2$ 
amplitude reduction factors should be
in a range that has been experimentally measured before, since this factor represents inelastic losses and errors in the theoretical backscattering amplitudes that only weakly depend on an individual system.  Indeed, our measurements fall
within acceptable ranges.\cite{Li95b,Booth98b} Second, 
the temperature dependence of the Debye-Waller factors
can be compared to a correlated-Debye model, with large offsets 
indicative of static (i.e. non-thermal) disorder or distortions.  
In all cases, we see no abnormally large offsets.
Third, 
the measured
pair distances should be reasonably close to those measured by diffraction,
which is consistent with our measurements (Table \ref{poly_table}).
Fourth,
various site-interchange possibilities should be considered, such as U sitting on
the (nominally) Sn (16c) site.  Although these fits are not particularly 
sensitive to such interchanges, our measurements are consistent with no site 
interchange.  Finally, the results from the single-crystal and the polycrystalline
samples are virtually identical.
Together with the single-crystal diffraction results,\cite{Shlyk99}
we must conclude 
that the U$_3$Ni$_3$Sn$_4$ system is structurally well ordered, and is
much more ordered than, say, the UPdCu$_4$ system.\cite{Booth98a,Bauer02}
Although such crystalline order does not rule out some other source of 
magnetic-interaction disorder that might be consistent with a Kondo disorder 
model or a Griffiths phase model, it certainly rules out extensive
lattice disorder.
In addition, although these structural studies cannot rule out the presence of 
small amounts of disorder, previous work within the simplest form of the
Kondo disorder model has 
shown that significantly more disorder would have to be present for that model 
to work (in UPdCu$_4$, for instance, Ref. \onlinecite{Bauer02} estimates that
at least 0.002 \AA$^2$ of static disorder is necessary to produce NFL behavior 
from the KDM).

Even with small, undetected amounts of disorder, Griffiths' phase models may be capable of describing the physics in this system.  In the AF-GP model, for 
instance, the exact relationship between the required degree of disorder as 
measured by the number and distribution of antiferromagnetic clusters and the 
physical properties remains unclear.  However, the AF-GP model makes very clear 
predictions about the evolution of the specific heat with applied magnetic
field.  We show in Sec. \ref{CHT} that these predictions do not describe the 
observed features.
Until more precise, measurable predictions relating disorder to physical
properties are obtained, we conclude that disorder presents at best a small
perturbation to this system.  Of course, some other specific-heat anomaly
of unknown origin may be responsible for the observed change in the FL/NFL 
crossover temperature with field.  However, without any possible candidates for 
such an anomaly, these results leave the AF-QCP theory as the only current
alternative.

There are other requirements for an AF-QCP theory, of course.  For one,
the system must be very near a magnetic/non-magnetic instability.  
Measurements under applied pressure indicate this instability may exist at a
small negative pressure, based on a scaling of the resistivity behavior.  We
thus expect that if the magnetic phase is antiferromagnetic, an applied magnetic
field will also move the system toward a Fermi liquid regime.  The data presented
in Fig. \ref{spec_fig} indicate this system behaves exactly as one
expects if the system is near a magnetic/non-magnetic instability, and
are therefore qualitatively consistent with models that include antiferromagnetism as a 
competing interaction.

\section{Conclusion}
\label{conclusion}

In summary, we have measured the local structure around the constituent atoms
in U$_3$Ni$_3$Sn$_4$ single crystals and polycrystals as a function of 
temperature and
stoichiometry.  These data follow typical Debye-model dependences in the 
measured pair-distance distribution widths with no static (i.e. non-thermal) 
offsets.  Moreover, the measured local structure agrees well with the previous
single-crystal diffraction studies.\cite{Shlyk99}  
In addition, the relatively low residual resistivity and all
other evidence indicates that this system is structurally well
ordered.  We also report specific heat data that are clearly inconsistent 
with the antiferromagnetic Griffiths' phase model.  These data instead suggest 
a recovery of 
Fermi liquid behavior under modest applied magnetic fields, qualitatively
consistent with the interpretation that
applied fields destroy magnetic fluctuations in the vicinity of a 
quantum critical point.  This result is also consistent with
work under applied pressure that indicates a negative-pressure 
QCP.\cite{Estrela01}
Taking all these results together, we have ruled out the simple form of the KDM 
and the AF-GP theories.  In addition, the lack of measurable disorder does not 
favor other disorder-based theories, such as the MIT-GP. 
With no indications to the contrary, we 
conclude that the best current description of U$_3$Ni$_3$Sn$_4$ is
that of a system near an antiferromagnetic quantum critical point.

\acknowledgments

We thank E. D. Bauer, A. J. Millis, E. Miranda, A. H. Castro Neto and D. L. Cox 
for many useful discussions.  This work was partially
supported by the Director, Office of Science, Office of Basic Energy Sciences 
(OBES),
Chemical Sciences, Geosciences and Biosciences Division, U.S. Department of
Energy (DOE) under Contract No. AC03-76SF00098.  Work at the University of 
Kentucky was supported by the DOE Office of Science, Division of Materials 
Sciences Grant No. DE-FG02-97ER45653.
XAFS data were collected at the Stanford Synchrotron Radiation
Laboratory, a national user facility operated by Stanford University
of the behalf of the DOE/OBES.

\bibliographystyle{apsrev}
\bibliography{/home/hahn/chbooth/papers/bib/bibli}

\begin{thebibliography}{40}
\bibitem[{u33()}]{u334_param}
\bibinfo{note}{A displacement parameter $x$=0.082 was obtained, but not
  reported, in Ref. \onlinecite{Shlyk99}.}
\end{thebibliography}

\end{document}